\documentclass[12pt]{article}
\usepackage{amssymb,psfrag,graphicx}
\title{Bounds on area and charge for marginally trapped surfaces
with cosmological constant \footnote{preprint UWThPh-2012-5}}
\author{}
\date{}
\begin{document}

\author{Walter Simon\\
Gravitationsphysik\\Fakult\"at f\"ur Physik, Universit\"at Wien\\
Boltzmanngasse 5,\\ 
A-1090 Wien, Austria} 

\maketitle

\begin{abstract}
We sharpen the known inequalities $A {\Lambda} \le 4\pi (1-g)$ \cite{HSN,EW} and $A \ge 4\pi Q^2$ \cite{DJR} between the area 
$A$ and the electric charge $Q$ of a stable marginally outer trapped surface
(MOTS) of genus $g$ in the presence of a cosmological constant $\Lambda$. 
In particular, instead of requiring stability we include the principal eigenvalue $\lambda$ of the stability operator. 
For $\Lambda^{*} =  \Lambda + \lambda > 0$ we obtain a lower and an upper bound for
$\Lambda^{*} A$ in terms of $\Lambda^{*}Q^2$, as well as the upper bound  $ Q \le
1/(2\sqrt{\Lambda^{*}})$ for the charge, which reduces to $ Q \le 1/(2\sqrt{\Lambda})$ 
in the stable case $\lambda \ge 0$. For $\Lambda^{*} < 0$ there remains only a lower bound on $A$.
In the spherically symmetric, static, stable case one of our area inequalities
is saturated iff the surface gravity vanishes. We also discuss implications of
our inequalities for "jumps" and mergers of charged MOTS.    
 \end{abstract}

We take marginally outer trapped surfaces (MOTS) to be smooth, connected, orientable
2-surfaces on which one of the orthogonally outgoing null geodesic congruences has vanishing
expansion. A key calculation in the theory of MOTS is the variation of the outgoing null expansion
in arbitrary directions. An application of this calculation are the arguments
leading to restrictions for the topology of MOTS when an energy condition is assumed (see e.g. \cite{HE}--\cite{GGA}). 
In the cosmological case, for either sign of $\Lambda$, Hayward, Shiromizu
and Nakao \cite{HSN} and Woolgar \cite{EW} refined this calculation to
obtain the bound 
 \begin{equation}
\label{A}
A {\Lambda} \le 4\pi (1-g)
\end{equation}
\cite{HSN,EW} where $A$ is the area and $g$ the genus of this surface.
On the other hand, in the presence of an electric field and for $\Lambda \ge 0$, Dain, Jaramillo and Reiris 
found the inequality
\begin{equation}
\label{Q}
4\pi Q^2 \le A 
\end{equation}
for the charge (Theorem 2.1 of \cite{DJR}), extending a previous observation by Gibbons
(formula (47) in \cite{GGI}) in the time-symmetric case.
This means that for $\Lambda > 0$ there is a lower as well as an upper bound for the area of a stable, 
charged MOTS, and hence the bound 
\begin{equation}
\label{Q<1}
Q \le \frac{1}{\sqrt{\Lambda}}
\end{equation} 
for the charge. 

By revisiting the original variational calculation we have obtained an
improvement of the above-mentioned inequalities which we formulate as
follows. \\  

\noindent
{\bf Theorem.} Let ${\cal S}$ be a MOTS in a spacetime $({\cal M}, g_{ab})$ where the Einstein
tensor $G_{ab}$  and the Maxwell and matter contributions to the energy momentum satisfy   
\begin{equation}
G_{ab} + \Lambda g_{ab} = T_{ab}^{Max} + T_{ab}^{mat} =  
 2\left(F_{ac}F^{~c}_{b} - \frac{1}{4} g_{ab} F_{cd}F^{cd} \right) 
+ T_{ab}^{mat} 
\end{equation}
and the latter satisfies the dominant enengy condition. 
Let $\lambda$ be the principal eigenvalue of the stability operator (\ref{st1}) defined in \cite{AMS1,AMS2}. 
Then in terms of $\Lambda^* = \Lambda + \lambda$, 
the electric charge $4\pi Q = \int_{\cal S} F_{ab}dS^{ab}$ and the rescaled
charge $q^* = 2 \sqrt{\Lambda^*} Q$  there hold the bounds 
(note that $q^*$ is imaginary when $\Lambda^* < 0$ !)
\begin{eqnarray}
\label{Q<2}
|q^*| & \le & 1  ~~ \mbox{for} ~ \Lambda^* > 0, \\
\label{A>}
2\pi \left[1  - \sqrt{1 - q^{*2}} \right]
& \le \Lambda^* A & \le 2\pi \left[ 1 + \sqrt{1 - q^{*2}}  \right]
 ~~ \mbox{for} ~ \Lambda^* > 0, \nonumber \\
{}\\
\label{A<}
2\pi \left[(g-1) + \sqrt{(g-1)^2 - q^{*2}} \right]& \le  - \Lambda^* A & {} 
\qquad \mbox{for} \quad \Lambda^* < 0. 
\end{eqnarray}
In the stable case $\lambda \ge 0$, all inequalities remain valid when
$\Lambda^{*}$ and $q^*$ are replaced by $\Lambda$ and  $q =
2\sqrt{\Lambda} Q$ respectively. 
\\

\noindent  
{\bf Proof.} We recall the terminology of \cite{DJR},\cite{AMS1}, and \cite{AMS2}:  
We denote by $l^a$ and $k^b$ the out- and ingoing null normals to ${\cal S}$,
and  by $\theta^{(l)}$, $\Omega_a^{(l)}$ and $\sigma_{ab}^{(l)}$ the expansion, the torsion and shear of
$l^a$, respectively.  
The covariant derivative $D_a$, the Laplacian $\Delta_{\cal S}$ and the scalar curvature $R_{\cal S}$ refer to ${\cal S}$.
We also recall the variation $\delta_{\psi v} \theta^{(l)}$ resp.
the stability operator $L_v(\psi)$ with respect to a normal direction $v^a = \beta l^a - k^a$, 
defined by (\ref{st1}), and rewritten in \cite{DJR} as (\ref{st2})
\begin{eqnarray}
\lefteqn{\psi^{-1} \delta_{\psi v} \theta^{(l)} =   \psi^{-1} L_v(\psi) =
- \psi^{-1}  \Delta_{\cal S} \psi +  D^a \Omega_a^{(l)} + 2 \psi^{-1} \Omega_a^{(l)} D_a \psi - {\Omega^{(l)}}^a \Omega_a^{(l)} + {} }\nonumber\\
\label{st1}
& & {} + \frac{1}{2} R_{\cal S} - \beta \left[ \sigma^{(l)}_{ab}{\sigma^{(l)}}^ {ab} + G_{ab}l^{a}l^b \right]
- G_{ab}k^a l^b = {} \\
& & {} =  - \Delta_{\cal S} \ln \psi + D^a \Omega_a^{(l)} - 
\left( D_a \ln \psi - \Omega_a^{(l)} \right)\left( D_a \ln \psi -
\Omega_a^{(l)} \right)
+ {} \nonumber\\
\label{st2}
& & {} + \frac{1}{2} R_{\cal S} - \beta \left[ \sigma^{(l)}_{ab}{\sigma^{(l)}}^ {ab} + G_{ab}l^{a}l^b \right]
- G_{ab}k^a l^b.
\end{eqnarray}
For any fixed $v^a$ this linear elliptic (but in general non self-adjoint) operator
acting on the function $\psi$ has a real "principal" eigenvalue (whose real part is lowest among all eigenvalues), 
and a corresponding real positive eigenfunction $\phi$, viz. $L_v\phi = \lambda \phi$. 
We now insert this eigenfunction and Einstein's equations in (\ref{st2}),
 integrate over ${\cal S}$ and perform the  manipulations of Theorem 2.1 of \cite{DJR} 
to estimate $\int_{\cal S} T_{ab}^{Max} dS^{ab}$ in terms of $Q^2/A$.  We obtain
\begin{equation}
\label{qu}
\Lambda^* A^2 - 4\pi(1-g) A + 16\pi^2Q^2 \le 0
\end{equation}    
which  yields the bounds  (\ref{Q<2}), (\ref{A>}) and (\ref{A<}).
The final assertion of the theorem holds since (\ref{qu}) remains valid if $\Lambda^{*}$ is replaced by $\Lambda$ for
$\lambda \ge 0$.\\

\noindent
{\bf Remarks.}
\begin{enumerate}
\item Comparing with \cite{HSN} and \cite{DJR}, the stability requirement on the MOTS has now been removed at the expense of introducing 
the principal eigenvalue  $\lambda$ of the stability operator. This can of course be done even in absence of cosmological or Maxwell terms.
\item A cosmological constant (of either sign) then just shifts the spectrum of the stability operator via
$\delta \rightarrow \delta^{*} = \Lambda + \delta$ for every eigenvalue $\delta$. 
\item 
The eigenvalue of the stability operator does not seem to have any direct physical
meaning. If there is any relation to quantum mechanical stability via
Hawking evaporation, it is an "inverse" one: For the bifurcate horizon sphere in
Schwarzschild with mass $M$, surface gravity $\kappa$ and temperature $T$, we
have $\lambda = 1/(4M^2) = 4\kappa^2 \propto T^2$. Hence "hot" black holes 
are "very stable" in our sense. 
We also note that (\ref{qu}) can be rewritten as an upper bound for $\lambda$.
\item
For $\Lambda^* > 0$ we obtain spherical topology $g=0$, 
which is a slight reformulation of the standard results on topology of MOTS \cite{GGA}. 
\item
Again for $\Lambda^* > 0$ the charge (\ref{Q<2}) which "fits into a black hole" is lower by a factor of
two compared to the estimate (\ref{Q<1}) following directly from (\ref{A}) and (\ref{Q}) in the stable case.
\item
The left half of (\ref{A>}) turns into (\ref{Q}) for  $q^{*2} \ll 1$ since $\sqrt{1 - q^{*2}}
\approx 1 - q^{*2}/2$, while (\ref{A<}) and the right half of (\ref{A>})
imply (\ref{A}) for
$q^* = 0$.
\item
In the spherically symmetric (hence static) case, it is easily checked that
the explicitly known ("Reissner-Nordstr\"om-deSitter-") solutions with stable MOTS
(cf Carter \cite{BC})  
saturate either one of (\ref{A>}) for $\Lambda^* > 0$ or (\ref{A<}) for $\Lambda^* < 0$ iff the surface gravity vanishes 
(in which case the MOTS in fact degenerates to the well known cylindrical end). Without the assumption of spherical
symmetry, the analogous result can be conjectured. 
\item
We can also introduce a generalized Christodoulou mass \cite{DC} and a generalized
surface gravity via the relations $M_g = M(A,Q,\Lambda)$
and $\kappa_g = \kappa(A,Q,\Lambda)$ satisfied by the spherically symmetric, static
solutions. Then the "first law of black hole mechanics" in the form 
\cite{IW} implies $\kappa_g = \partial M_g/\partial A$.
For the general (dynamical) case in which the area increases monotonically
 with time,  this yields monotonicity properties for 
 $M_g$ depending on the sign of $\kappa_g$ as exposed in \cite{AK} and
 \cite{SD}.
\item  Magnetic and Yang-Mills charges can be included straightforwardly 
in the above bounds, (compare \cite{DJR,JJ})
\item In the axially symmetric case, the cosmological constant would be a
highly interesting and non-trivial addendum to the bound
\begin{equation}
\label{AJQ}
\frac{A^2}{16\pi^2} \ge 4J^2 + Q^2
\end{equation}
for the area in terms of the
angular momentum $J$ and $Q$ obtained in \cite{JRD,GJ} when $\Lambda^* \ge 0$. 
In addition to $q^* = 2 \sqrt{\Lambda^*} Q$, 
 the parameter $j^{*} = \Lambda^* J$ will play a crucial role in this case
as can already be seen from combining (\ref{AJQ}) with (\ref{A}) or (\ref{A>}). Here the natural task is, however, to obtain bounds on $A$,
$q^*$ and $j^*$ which are sharp in the extreme cases (cf. remark 7). This will be described elsewhere.
 \end{enumerate}

As an application of our bounds we consider a spacetime foliated by spacelike hypersurfaces, 
with a (possibly disconnected) trapped region and an untrapped outer barrier on each leaf. Then as
shown by Andersson and Metzger \cite{AM} and Eichmair \cite{ME} each such leaf
contains unique, smooth, and stable "outermost MOTS" (possibly consisting of
several components). 
Upon time evolution, MOTS are capable of "jumping"  \cite{AMMS} (which includes "merging"
if there are several MOTS present initially). 
Accordingly, on a slice corresponding to a "jump time" there are at least two homologous
MOTS.
However, electric charges as well as a positive cosmological constant provide repulsive 
forces which loosely speaking restrict the options for jumping or forbid jumps at all.
Consider in particular a foliated spactime with $\Lambda > 0$ and an
untrapped barrier on each slice. Then two MOTS with charges $Q_1$ and $Q_2$ such that
$Q_1 + Q_2 > 1/(2\sqrt{\Lambda})$ cannot merge on such a spacetime,
irrespective of their stability, 
as the stable outermost target of the jump would violate (\ref{Q<2}).  
   
We note that a restriction for the merging of charged black holes for
$\Lambda > 0$ 
was also  obtained by Shiromizu, Nakao, Kodama and Maeda \cite{SNKM}. However, this
result is based on the existence of an event horizon and an area law for it, which
hinges on several assumptions. On the other hand, in the present setting 
a general area law for MOTS under jumps is unknown, except that 
 the initial and the final MOTS have to respect the area bounds 
(\ref{A>}) and (\ref{A<}). For stable MOTS this restriction can be formulated as 
follows.
\\

\noindent
{\bf Corollary.} We consider a spacelike hypersurface ${\cal N}$ in the
spacetime ${\cal M}$ satisfying the requirements of the Theorem with $\Lambda >0$.
\begin{enumerate}
\item
Let ${\cal N}$ contain two homologous, stable MOTS ${\cal S}_i$, ${\cal S}_f$ with
areas $A_i$ and $A_f$ such that the domain bounded by ${\cal S}_i$, ${\cal S}_f$
is free of charges. Then for $\Delta A = A_i/A_f$ and in terms of the rescaled total charge  
$q$ we find 
\begin{eqnarray}
\label{dA1}
 \frac{1 - \sqrt{1 - q^{2}}} 
{1 + \sqrt{1 - q^{2}}} \le 
\Delta A  \le  \frac{1 + \sqrt{1 - q^{2}}}
{1 - \sqrt{1 - q^{2}}} 
\end{eqnarray}
\item
Let ${\cal N}$ contain three stable MOTS ${\cal S}_{i,1}$, ${\cal S}_{i,2}$ and ${\cal S}_f$ 
with areas $A_{i,1}$ and $A_{i,2}$  and $A_f$,
${\cal S}_{i,1} \cup {\cal S}_{i,2}$ homologous to ${\cal S}_f$, 
and the intermediate domain free of charges.
Let  $q_1$, $q_2$ be the charges of ${\cal S}_{i,1}$, ${\cal S}_{i,2}$ such
that $q_1 q_2 > 0$. Then for $\Delta A = (A_{1,i} + A_{2,i})/A_f$ we obtain  
\begin{eqnarray}
\label{dA2}
\frac{2 - \sqrt{1 - q_1^{2}} -  \sqrt{1 - q_2^{2}}}
{1 + \sqrt{1 -  (q_1^{2} + q_2^{2})}} \le 
\Delta A  \le  \frac{2 + \sqrt{1 - q_1^{2}} +  \sqrt{1 - q_2^{2}}}
{1 - \sqrt{1 - (q_1^{2} + q_2^{2})}} 
\end{eqnarray}

\end{enumerate}

\noindent
{\bf Proof.} (\ref{dA1}) is obvious from (\ref{A>}) in which 
$q^*$ can be replaced by $q$ due to stability. 
To get (\ref{dA2}) we have used  charge conservation $q = q_1 + q_2$ 
for homologous surfaces, as well as $q^{2} \ge q_1^{2} +  q_2^{2}$ for
charges of equal sign.
\\

\noindent
{\bf Remarks}. 
\begin{enumerate}
\item A generalization to more MOTS-components is trivial. On the other
hand, upon relaxing the stability conditions the explicit appearance of 
$\lambda$ in the area bounds seems inevitable.   
\item
Compared to (\ref{dA1}) a simpler but rougher bound on $\Delta A$
(under the same requirements), reads $\left(4q^{-2} -1 \right)^{-1} \le
\Delta A \le 4 q^{-2} -1$. This is obvious from (\ref{dA1}) and
 $\sqrt{1 - q^2} < 1 - q^2/2$. (\ref{dA2}) can be simplified in
 the same manner. 
\end{enumerate}

\section*{Acknowledgement}
I am grateful to Lars Andersson, Sergio Dain, Jos\'e-Luis Jaramillo, Mart\'in Reiris
and to a referee for helpful comments.   
This work was supported by the Austrian Science Fund (FWF) P23337-N16.

\end{document}